# Evaluation of 50 Greek Science and Engineering University Departments using Google Scholar

Marina Pitsolanti, Fotini Papadopoulou, Nikolaos Tselios

Department of Educational Sciences and Early Childhood Education, University of Patras, University Campus, 26500 Rio, Patras, Greece

mpitsolanti@gmail.com, fotini1987@gmail.com, nitse@ece.upatras.gr

**Abstract**. In this paper, the scientometric evaluation of faculty members of 50 Greek Science and Engineering University Departments is presented. 1978 academics were examined in total. The number of papers, citations, h-index and i10-index have been collected for each academic, department, school and university using Google Scholar and the citations analysis program Publish or Perish. Analysis of the collected data showed that departments of the same academic discipline are characterized by significant differences on the scientific outcome. In addition, in the majority of the evaluated departments a significant difference in h-index between academics who report scientific activity on the departments' website and those who do not was observed. Moreover, academics who earned their PhD title in the USA demonstrate higher indices in comparison to scholars who obtained their PhD title in Europe or in Greece. Finally, the correlation between the academic rank and the scholars' h-index (or the number of their citations) is quite low in some departments, which, under specific circumstances, could be an indication of the lack of meritocracy.

## 1. Introduction

Scientometric evaluation and ranking of universities, departments and scholar constitutes a widely accepted topic which informs decisions both for students as well as for academics. In general, two main factors are subject to evaluation: The quality of the educational process offered by a specific institute or scholar and the quality and the quantity of the scientific outcome. The latter is largely based on scientific indices such as as the total number of publications, citations, h-index, i-10 index and others, such as m-index, which are subsequently calculated using the before mentioned indices (Hirsch,

2005). The process of collecting such data is greatly facilitated by web-based scientific databases, namely Google Scholar, Web of Science and Scopus. In addition, there are other databases such as EconLib and PubMed, which focus on a specific scientific discipline.

One of the most popular indicators for scholars' scientific quality evaluation is the h index Hirsch (2005). The main advantage of the index is that it combines both scientific productivity and quality as measured by the impact of his work to the rest of the scientific community in terms of citations (Glänzel, 2006). The h index is based on the distribution of citations of a specific scholar and is calculated as follows: a researcher has h-index equal to n if she has n publications with at least n references each and every other publication received less than n citations (Hirsch, 2005).

A second advantage of h index is that it is quite easy to evaluate objectively and quantitatively a researcher, a process which is necessary for decisions related to professors' hiring and promotion, research funding and for nominating awards such as the Nobel and Turing prize (Hirsch, 2005). For instance, Hirsch (2005) calculated the h-index of Nobel Prize winners and found that 84% of them have exceeds 30. Moreover, newly elected members of the US National Academy of Sciences of physics and Astronomy of 2005 had an average h-index 46 at the time.

At the same time, by using the h index the comparison between researchers is accomplished in a more complete manner (Hirsch, 2005). Not suprisingly h-index is used many times as a measurement instrument of scientists, journal and departments of different disciplines (Abramo, Cicero, & D'Angelo, 2012, Altanopoulou, Tselios, & Dontsidou, 2012, Egghe, 2010, Glänzel, Schubert, & Lazaridis, 2010, Kazakis, 2014, Kazakis, Diamantidis, Fragidis, & Lazarides, 2014, Shin, Toutkoushian, & Teichler, 2011, Thijs, & Debackere, K., 2011). Moreover, during university rankings of computer science departments, it was found that the rankings based on measurement reports for a scholar and their ratings according to the h index, showed a strong positive correlation (Cronin & Meho, 2006). Thus, nowadays it is widely accepted that the h-index offers a meaningful way to identify differences in scholars, departments or journals (Cronin & Meho, 2006, Glanzel, 2006, Mingers, & Leydesdorff, 2015).

Beyond the obvious advantages of h-index, there are some drawbacks (Mingers, & Leydesdorff, 2015, Moed, 2006). The h index grows with new publications which can receive a number of citations, but it also grows with new references added to already existing articles. So a researcher can increase her h-index, without having to publish for a time (Glänzel, 2006). Another drawback is that by using h for measuring the quality of a younger scientist the results are not always representative. That is because a new researcher is not active for the same period of time, and time is required to receive citations in new articles. This is especially true in the Social sciences and Humanities, an article can take five to ten years to receive a significant number of citations (Harzing, 2010, p. 10). Another shortcoming of the h-index is that it doesn't take into account any attention to the number of writers who participate in a paper as well as the contribution of each one in this (Hirsch, 2005). Thus, a researcher can increase the h index through collaborations where their role is not significant, apart from contributions where he is the main or the second author.

Apparently, such an approach could lead to superficial results since the publication practices as well the mean impact factor of each field vary significantly (Costas & Bordons, 2007). Differences in standard h values in the various fields, are mainly influenced by the average number of citations to a paper, the average number of publications produced by every scientist in the field, and the size (number of scientists) in the field (Egghe, 2010). Therefore, it would be better if the comparison is carried out strictly between researchers belonging to the same scientific field or at least normalize the results in order to be comparable. For instance, Batista et al. (2006) report that the ratio between the mean h indices for the scientific disciplines of biology and mathematics is 3:1. Scientists working in 'smaller' or marginal scientific areas will not achieve the same high h values, compared with those who work in extremely topical areas (Hirsch, 2005).

*Study objectives and questions*

In general, this paper aims to highlight the positive contribution of scientometrics and its usefulness in issues related to higher education quality evaluation (Marchant, & Bouyssou, 2011). In specific, it attempts to provide answers to the following research questions:

1. Are there any significant differences in publications, citations, h index and i10 index between departments of the same discipline which are located in different universities?
2. Are there any differences on the academics' scientific performance (as expressed using the h-index) between those who report detailed information about their research on the department's website and those who don't?
3. Are there any differences on the scholars' performance according to the location in which they completed their PhD (namely Greece, other European countries, or USA)?
4. Is there any correlation between the academics' rank (Full Professor, Associate Professor, Assistant Professor, Lecturer) and their h- index and total number of citations?

Correlation (correlation) and (a) h-index (b) citations/references

The rest of the paper is organized as follows. First, the research design is described in detail and the tools used to collect the data are discussed in brief. Subsequently, the results of the analysis are presented for each department and for each research question. Finally, the obtained findings are discussed and future goals are derived.

## 2. Methodology

*Research Design*

Fifty departments from Science and Engineering disciplines were selected for the study (see Table 1). All in all, 31 Science and 19 Engineering departments were evaluated. The procedure proposed by Altanopoulou, Dontsidou and Tselios (2012) was followed to record and analyze the data. The names, surnames and academic grade of all the academics were recorded. The program Publish or Perish (PoP) was used to calculate the total publications, citations, h-index, i-10 index, m-index. If a scholar used at the time Google Scholar Profile the related data were collected from there, instead of using PoP.

Table 1. Evaluated departments for each scientific field

| Field | Number of evaluated departments | Department's name | Number of academics of each department |
|---|---|---|---|
| Natural & Information sciences | *31* | Mathematics (6) | 206 |
| | | Statistics (2) | 48 |
| | | Physics (5) | 270 |
| | | Biology (5) | 187 |
| | | Chemistry (5) | 245 |
| | | Informatics (8) | 189 |
| Technological sciences | 19 | Civil engineering (5) | 259 |
| | | Chemical engineering (3) | 134 |
| | | Mechanical engineering (5) | 156 |
| | | Electrical & computer engineering (6) | 284 |
| **Total** | **50** | | **1978** |

Subsequently, for each department the following indices were calculated: the mean median and standard deviation on publications, citations, h-index, i-10 index and the mean m-index, the percentage of academic members who report information on their website and the percentage of academic members who retain a Google Scholar Profile. Subsequently, the aggregate results were calculated for each department, as well as for the departments of the same discipline. In some cases, synonymy could slightly affect the presented data, since there is always the possibility of having two scholars with the same name and surname. In such cases, a Google Scholar profile greatly assisted the procedure. If this was not the case the data were cleaned and the affiliation of each author was closely examined. However, it is difficult to claim 100% success while in the process of evaluating 1978 faculty members (Altanopoulou, Dontsidou, &Tselios 2012).

As mentioned above, the Google Scholar database was the source to retrieve the scientometric data. In addition to free access offered by Google Scholar, there are still 3 advantages which characterize its use. Google Scholar is easy and straightforward to use. It is also quite efficient, because the search of information takes place immediately without needing additional registration steps to access the available data. Finally, the main advantage of Google Scholar is the wide coverage of scientific disciplines and publication venues which surpasses both Scopus and Web of Science (Bar-Ilan, 2008b). The information related to the scientific activity of a specific scholar, covers not only notable and reputable scientific journals, but it also contains references from books or book chapters, conference proceedings and technical reports reports which are not indexed in Web of Science and Scopus databases (Harzing, 2010, Mingers, & Leydesdorff, 2015).

As far as the collection of data is concerned each of the databases follows a different approach, which in turn affects the results in total numbers of publications and reports (Bar-Ilan, 2008a, Bar-Ilan, 2008b). For this reason, and in conjunction with the margin of error in the algorithm of Google Scholar, there is the possibility that the number of citations for a specific to is actually smaller (or even higher) than it appears. This can happen for a variety of reasons such as unrecognized text format, or error in recognition of the date of publication (Jacso, 2006). In the process of scientometric evaluation, various indicators are nowadays widely accepted and used such as the total number of publications, citations, h-index, i-10 index and others, such as m-index, which are calculated using the before mentioned indices (Hirsch, 2005). The data for all faculty members and department were collected from 10 March 2015 to June 1st, 2015. The data were recorded and analyzed using Google Sheets and SPSS v.21 and are presented in the following section.

## 3. Presentation of the results

In this section, the aggregate evaluation results are presented by subject area. The following data are presented: the number of academics in a department; median and mean number or publications per faculty member (as well as standard deviation); median and mean number of citations per academic (and standard deviation); median and mean number of h-index (and standard deviation) mean and median number of i10-index (number of papers which have at least 10 citations each). Moreover, percentage of academics who report scientific activity on the departments' web site and the percentage of scholars who retain Google Scholar Profile is presented. The departments were ranked according to their median h-index.

**RQ1. Variation between departments of the same scientific discipline.**

*3.1 Departments of Science*

In the departments of *Mathematics* (Table 1), although the department at the University of Crete precedes in terms of academics' median h, the department at the University of Ioannina scores highest in mean (and median) number of publications, median number of citations and mean h-index (Table 1). The department of Athens has, by far, the most members. Among all departments, the departments at the University of Crete and Ioannina have the highest percentage of academics who report scientific activity and their publications in the departments' website.

Table 1: Aggregate results of Departments of Mathematics

| DEPARTMENTS OF MATHEMATICS | | | | | | | | | | | | | | m-index | Res. Act. | GS pr. |
|---|---|---|---|---|---|---|---|---|---|---|---|---|---|---|---|---|
| | | Publications | | | Citations | | | h - index | | | i10-index | | | | | |
| University | No. | Mean | SD | Median | Mean | SD | Median | Mean | SD | Median | Mean | SD | Median | Mean | | |
| Crete | 28 | 31.3 | 13.2 | 29.5 | 359.3 | 318 | 279 | 8.8 | 3.7 | **9.5** | 8.6 | 6.2 | **9.5** | 0.4 | **95.7%** | 14.3% |
| Ioannina | 23 | **50.5** | 46 | **40** | 519.5 | 573.9 | **307** | **10.1** | 6.9 | 9 | **12.7** | 13.9 | 8 | **0.5** | **95.7%** | 21.7% |
| Athens | 55 | 47.3 | 46 | 32 | 473.9 | 678.5 | 224 | 9.4 | 6.5 | 8 | 11.8 | 15 | 7 | 0.4 | 41.8% | 23.6% |
| Patra | 38 | 48.4 | 65 | 29 | **669.9** | 1724.4 | 192 | 9.2 | 8.5 | 7.5 | 11.6 | 25.6 | 6 | 0.4 | 81.6% | **23.7%** |
| Thessaloniki | 25 | 39.2 | 25.1 | 34 | 269 | 326.3 | 174 | 7.3 | 4.1 | 7 | 6.2 | 6 | 5 | 0.3 | 56% | 16% |
| Aegean | 37 | 34.8 | 26.7 | 32 | 209.3 | 200.8 | 149 | 6.1 | 3.4 | 6 | 5.5 | 4.6 | 5 | 0.4 | 70.3% | 16.2% |
| **Mean** | *34.3* | *41.9* | *37* | *32.8* | *416.8* | *637* | *220.8* | *8.5* | *5.5* | *7.8* | *9.4* | *11.9* | *6.8* | *0.4* | *73.5%* | *19.3%* |
| **SD** | *10.8* | *7.3* | *43.1* | *3.6* | *156.1* | *875.6* | *56.3* | *1.3* | *6.1* | *1.2* | *2.8* | *14.9* | *1.6* | *0.1* | *19.9%* | *3.9%* |
| **Median** | *32.5* | *43.3* | *36.3* | *32* | *416.6* | *450.1* | *208* | *9* | *5.3* | *7.8* | *10.1* | *10* | *6.5* | *0.4* | *75.9%* | *19%* |

Notes. No.: number of academics serving in each department; Res. Act.: percentage of academics who report scientific activity on the department's website; , GS Pr= % of academics maintaining Google Scholar Profile; Publications: lifetime Google Scholar's publications per academic (standard deviation and median); Citations: Citations per academic; Mean h: total of academics' h-index subsequently divided by the total of academics; i10-

index: number of papers which have at least 10 citations each, per academic; Figures in bold font indicate the highest value in each column.

One may notice significant differences in mean and median number of publications, citations and h-index between the departments. This demonstrates that departments in the same scientific subject, which have the same resources (for example, financial support from the Ministry of Education, comparable infrastructure and exactly the same wage for each academic according to their grades), have notable differences in research outcomes. However, no official national report states those differences. This is also evident in other scientific disciplines, as discussed below.

Table 2: Aggregate results of departments of Statistics

| | | \multicolumn{3}{c|}{Publications} | \multicolumn{3}{c|}{Citations} | \multicolumn{3}{c|}{h - index} | \multicolumn{3}{c|}{i10-index} | m-index | Res. | GS pr. |
|---|---|---|---|---|---|---|---|---|---|---|---|---|---|---|---|---|
| University | No. | Mean | SD | Median | Mean | SD | Median | Mean | SD | Median | Mean | SD | Median | Mean | Act. | |
| Athens(OPA) | 24 | **56.6** | 44.4 | 36.5 | **664.1** | 599.1 | **468.5** | **11.1** | 5.5 | **11** | **14** | 11.7 | **11.5** | 0.6 | **87,5%** | **66,7%** |
| Piraeus | 24 | 41.2 | 25.2 | 37.5 | 362 | 460.6 | 176.5 | 8.6 | 4.9 | 8 | 9.3 | 10.6 | 6 | 0.5 | **87,5%** | 33,3% |
| Mean | 24 | 48.9 | 34.8 | 37 | 513.1 | 529.9 | 322.5 | 9.9 | 5.2 | 9.5 | 11.7 | 11.1 | 8.8 | 0.5 | 87,5% | 50% |
| SD | 0 | 7.7 | 36.9 | 0.5 | 151.1 | 555.3 | 146 | 1.3 | 5.3 | 1.5 | 2.4 | 11.4 | 2.8 | 0 | 0,0% | 16,7% |
| Median | 24 | 48.9 | 34.8 | 37 | 513.1 | 529.9 | 322.5 | 9.9 | 5.2 | 9.5 | 11.7 | 11.1 | 8.8 | 0.5 | 87,5% | 50% |

Table 2 presents the scientometric data for the two Greek departments of *Statistics*. First comes the department of Athens on all indices expect the median number of publications. Faculty members in both departments have the same percentage of academics who report scientific activity and their publications in the departments' website (87.5%). However, the department of Athens shows best results at far maintenance of Google Scholar profiles is concerned.

Table 3: Aggregate results of departments of Physics

| | | \multicolumn{3}{c|}{Publications} | \multicolumn{3}{c|}{Citations} | \multicolumn{3}{c|}{h - index} | \multicolumn{3}{c|}{i10-index} | m-index | Res. | GS |
|---|---|---|---|---|---|---|---|---|---|---|---|---|---|---|---|---|
| University | No. | Mean | S D | Median | Mean | S D | Median | Mean | S D | Median | Mean | S D | Median | Mean | Act. | pr. |

| University | No. | Mean | S D | Median | Mean | S D | Median | Mean | S D | Median | Mean | S D | Median | Mean | Res. Act. | GS pr. |
|---|---|---|---|---|---|---|---|---|---|---|---|---|---|---|---|---|
| Crete | 25 | **130.7** | 87.1 | 86 | **2871.3** | 2777 | **1985** | **25.5** | 12.1 | **23** | **54.9** | 40.4 | **44** | 1.1 | **100%** | 20% |
| Ioannina | 51 | 92.2 | 58.5 | 85 | 1518.3 | 1556.2 | 1025 | 17.4 | 10 | 17 | 30.1 | 26.3 | 24 | 0.9 | 13.7% | 13.7% |
| Athens | 78 | 122.8 | 131.5 | 96.5 | 1745.3 | 2513.9 | 819 | 18.4 | 11.8 | 16.5 | 34.8 | 42.2 | 23.5 | 0.7 | 56.4% | 23.1% |
| Thessaloniki | 79 | 127 | 78.6 | **108** | 1706.2 | 1907.1 | 1092 | 18.4 | 9.2 | 16 | 35 | 30.1 | 27 | 0.8 | 87.3% | 34.2% |
| Patra | 37 | 63.7 | 34.6 | 60 | 835 | 763.2 | 639 | 14.2 | 6.5 | 13 | 21.3 | 15.6 | 20 | 0.9 | 83.8% | **37.8%** |
| **Mean** | 54 | 107.3 | 78.1 | 87.1 | 1735.2 | 1903.5 | 1112 | 18.8 | 9.9 | 17.1 | 35.2 | 30.9 | 27.7 | 0.9 | 68.3% | 25.8% |
| **SD** | 21.6 | 25.7 | 94 | 15.9 | 655.6 | 2092.1 | 464.7 | 3.7 | 10.5 | 3.3 | 11 | 34.1 | 8.4 | 0.1 | 30.7% | 9% |
| **Median** | 51 | 122.8 | 78.6 | 86 | 1706.2 | 1907.1 | 1025 | 18.4 | 10 | 16.5 | 34.8 | 30.1 | 24 | 0.9 | 83.8% | 23.1% |

Regarding the departments of *Physics* (Table 3) it appears that the department of Crete, has by far the highest scores in most evaluation indices of the scholars' research work, with a significant difference from the other parts. A notable result is the great difference on the mean h index between the department of Crete and the department at the University of Patras which has been ranked last (25.5-14.2). The department of Ioannina has the lowest scores in terms of faculty members' web site reported activity and Google Scholar Profile (13.7%). In the Department of Crete, while 100% of the faculty members report their scientific activity on the department's web site, only 20% of them have a Google Scholar profile.

Table 4: Aggregate results of departments of Chemistry

| DEPARTMENTS OF CHEMISTRY | | | | | | | | | | | | | | | | |
|---|---|---|---|---|---|---|---|---|---|---|---|---|---|---|---|---|
| | | Publications | | | Citations | | | h - index | | | i10-index | | | m-index | | |
| University | No. | Mean | S D | Median | Mean | S D | Median | Mean | S D | Median | Mean | S D | Median | Mean | Res. Act. | GS pr. |
| Crete | 25 | **127.9** | 73.4 | **111** | **3146.7** | 2489.5 | **2740** | **28** | 9.7 | **26** | **59.4** | 32.5 | **51** | **1.2** | 72% | 12% |
| Patra | 37 | 108.2 | 85.2 | 81 | 1835.3 | 1805 | 1149 | 20.6 | 9.6 | 19 | 45.4 | 43.7 | 30 | 0.8 | **89.2%** | 24.3% |
| Thessaloniki | 82 | 86 | 57 | 69 | 1646.1 | 1638.7 | 1051 | 19.8 | 9.9 | 18 | 36 | 28.9 | 27 | 0.7 | 64.6% | 12.2% |
| Ioannina | 50 | 76.9 | 44.3 | 71 | 1475.6 | 1600.5 | 970 | 18.7 | 8.2 | 18 | 31.8 | 21.3 | 27 | 0.8 | 82% | 10% |
| Athens | 51 | 84.1 | 66.5 | 56 | 1388.8 | 1387.8 | 888 | 17.5 | 8.2 | 17 | 31.7 | 24.9 | 24 | 0.8 | 86.3% | **100%** |
| **Mean** | 49 | 96.6 | 65.3 | 77.6 | 1898.5 | 1784.3 | 1359.6 | 20.9 | 9.1 | 19.6 | 40.9 | 30.3 | 31.8 | 0.9 | 78.8% | 31.7% |
| **SD** | 19 | 18.8 | 65.6 | 18.5 | 642.5 | 1787.5 | 695.6 | 3.7 | 9.6 | 3.3 | 10.6 | 31.2 | 9.8 | 0.1 | 9.2% | 34.5% |
| **Median** | 50 | 86 | 66.5 | 71 | 1646.1 | 1638.7 | 1051 | 19.8 | 9.6 | 18 | 36 | 28.9 | 27 | 0.8 | 82% | 12.2% |

Among the departments of *Chemistry* (Table 4), the department at the University of Crete is ranked first in all indices: h-index, publications, citations, i10-index, m index. The department of Athens has the lowest indexes (citations, i10-index, h-index), but 100% of its members maintain Google Scholar profile. The department at the University of Ioannina has the lowest mean number of publications.

Table 5: Aggregate results of departments of Biology

| | | Publications | | | Citations | | | h - index | | | i10-index | | | m-index | | |
|---|---|---|---|---|---|---|---|---|---|---|---|---|---|---|---|---|
| **DEPARTMENTS OF BIOLOGY** | | | | | | | | | | | | | | | | |
| University | No. | Mean | SD | Median | Mean | SD | Median | Mean | SD | Median | Mean | SD | Median | Mean | Res. Act. | GS pr. |
| Crete | 27 | **78.1** | 46.7 | **66** | **1810.4** | 832.8 | **1861** | **20.6** | 6.8 | **22** | **33.5** | 18.1 | **32** | **0.92** | **100%** | 25.9% |
| Thessaloniki | 59 | 70.4 | 40.4 | 63 | 1198.5 | 1152.6 | 885 | 16.3 | 6.3 | 16 | 25.7 | 16.3 | 21 | 0.7 | 89.8% | 23.7% |
| Athens | 49 | 65.4 | 42 | 50 | 1134.6 | 1094.6 | 833 | 16.2 | 7 | 15 | 43.6 | 138.8 | 22 | 0.69 | 98% | 24.5% |
| Patra | 31 | 50.5 | 27.1 | 43 | 834 | 829.4 | 627 | 14.4 | 6.6 | 14 | 20.4 | 15.3 | 19 | 0.68 | 100% | 29% |
| Thrace | 21 | 39.7 | 23.6 | 39 | 1080.4 | 843.6 | 838 | 13 | 5.9 | 14 | 16.8 | 10.1 | 15 | 0.73 | 81% | 14.3% |
| **Mean** | *187* | *60.8* | *36* | *52.2* | *1211.6* | *950.6* | *1008.8* | *16.1* | *6.5* | *16.2* | *28* | *39.7* | *21.8* | *0.74* | *93.7%* | *23.5%* |
| **SD** | *14.3* | *13.9* | *40.2* | *10.7* | *323.9* | *1051.6* | *435.3* | *2.6* | *6.9* | *3* | *9.6* | *73.1* | *5.6* | *0.09* | *7.4%* | *4.9%* |
| **Median** | *31* | *65.4* | *40.4* | *50* | *1134.6* | *843.6* | *838* | *16.2* | *6.6* | *15* | *25.7* | *16.3* | *21* | *0.7* | *98%* | *24.5%* |

According to the above mentioned results (Table 5), in the departments of *Biology*, the department at the University of Crete surpasses all the others in all indices. On the other hand, as far as h-index is concerned all the other departments have a comparable median number. On the other hand, the newly established department at the University of Thrace has relatively small scores in publications, h-index, i10-index, percentage of scholars who report scientific activity on the Internet and who maintain a Google Scholar profile.

Table 6: Aggregate results of Computer Science departments

| | | Publications | | | Citations | | | h - index | | | i10-index | | | m-index | | |
|---|---|---|---|---|---|---|---|---|---|---|---|---|---|---|---|---|
| **DEPARTMENTS OF COMPUTER SCIENCE** | | | | | | | | | | | | | | | | |
| University | No. | Mean | S D | Median | Mean | S D | Median | Mean | S D | Median | Mean | S D | Median | Mean | Res. Act. | GS pr. |

| University | No. | Publications Mean | SD | Median | Citations Mean | SD | Median | h-index Mean | SD | Median | i10-index Mean | SD | Median | m-index Mean | Res. Act. | GS pr. |
|---|---|---|---|---|---|---|---|---|---|---|---|---|---|---|---|---|
| Athens(UOA) | 42 | 144.9 | 82.4 | **136.5** | 2325.1 | 2733.7 | **1498** | 21 | 11.5 | **21** | 40 | 40 | 32 | 0.94 | 90.5% | 35.7% |
| Thessaloniki | 28 | **166.8** | 152.7 | 121.5 | **2565.7** | 3945.8 | 1258 | **21.1** | 12.6 | 18.5 | **47.9** | 54.1 | **39** | 1.19 | **96.4%** | **75%** |
| Athens(AUEB) | 33 | 96.5 | 71.9 | 63 | 2004.9 | 2587.7 | 783 | 17.2 | 10 | 15 | 30.1 | 26.9 | 24 | 0.94 | 78.8% | 66.7% |
| Thessaly | 11 | 72.5 | 40.1 | 82 | 1042.6 | 698.6 | 875 | 15.1 | 7.2 | 15 | 23.3 | 15.8 | 18 | 1.23 | 63.6% | 72.7% |
| Piraeus | 22 | 109.2 | 97.8 | 72 | 1233 | 1625.2 | 706 | 14.4 | 9.7 | 14 | 24 | 24.6 | 16.5 | 0.71 | 68.2% | 27.3% |
| Tripoli | 26 | 112 | 152.5 | 91 | 1176.1 | 1804.4 | 650.5 | 15.4 | 9 | 14 | 29.5 | 46.1 | 18 | 1.02 | 88.5% | 73.1% |
| Athens (Harokopio) | 11 | 76.3 | 58.4 | 49 | 474 | 309.9 | 431 | 10.4 | 4 | 11 | 13.9 | 9.2 | 15 | 0.93 | 90.9% | 27.3% |
| Ionian | 16 | 66.9 | 32.3 | 59 | 560.1 | 354.5 | 528.5 | 11.3 | 4 | 11 | 11.7 | 8.1 | 13 | 0.97 | 100% | 50% |
| **Mean** | 23.6 | 105.6 | 86 | 84.3 | 1422.7 | 1757.5 | 841.3 | 15.7 | 8.5 | 14.9 | 27.5 | 28.1 | 21.9 | 0.99 | 84.6% | 53.5% |
| **SD** | 10.2 | 33.3 | 107.7 | 28.8 | 737 | 2537 | 341 | 3.7 | 10.5 | 3.2 | 11.4 | 35.3 | 8.6 | 0.15 | 12.3% | 19.7% |
| **Median** | 24 | 102.9 | 77.1 | 77 | 1204.5 | 1714.8 | 744.5 | 15.2 | 9.4 | 14.5 | 26.8 | 25.7 | 18 | 0.96 | 89.5% | 58.3% |

At Table 6, it appears that the *Computer Science* departments which show the highest scores in the evaluation indices are the department at the Universities of Athens and Thessaloniki. As far as mean h-index is concerned, third is ranked another department located in Athens but at the Athens University of Economics and Business (AUEB). On the other hand, the departments at the Harokopio and Ionian University have the lowest scores on publications, citations, h-index and, i10-index. In general, the online reporting rates are quite high for all departments, as well as the degree of adoption of GS profile among the scholars.

*3.2 Departments of Engineering*

Table 7: Aggregate results of Chemical Engineering departments

| DEPARTMENTS OF CHEMICAL ENGINEERING ||||||||||||||||
|---|---|---|---|---|---|---|---|---|---|---|---|---|---|---|---|---|
| | | Publications ||| Citations ||| h - index ||| i10-index ||| m-index | Res. Act. | GS pr. |
| University | No. | Mean | SD | Median | Mean | SD | Median | Mean | SD | Median | Mean | SD | Median | Mean | | |
| Patra | 31 | **115.2** | 89.6 | **86** | **3595.4** | 5814.3 | 1502 | **25.7** | 16 | **22** | **54.5** | 48.9 | **40** | 1.1 | **100%** | 64.5% |
| Thessaloniki | 34 | 114.1 | 99.3 | 83.5 | 1943.4 | 1859 | **1592** | 21.1 | 10.7 | **22** | 39.8 | 37.3 | 33.5 | 0.9 | **100%** | 26.5% |
| NTUA | 69 | 94.2 | 62.4 | 84 | 1710.2 | 1714.4 | 1298 | 19.8 | 10.6 | 18 | 34.5 | 28.7 | 25 | 0.8 | 78.3% | 34.8% |
| **Mean** | 44.7 | 107.8 | 83.8 | 84.5 | 2416.3 | 3129.2 | 1464 | 22.2 | 12.5 | 20.7 | 42.9 | 38.3 | 32.8 | 0.9 | 92.8% | 41.9% |
| **SD** | 17.2 | 9.7 | 80.4 | 1.1 | 839.1 | 3286.6 | 123 | 2.6 | 12.3 | 1.9 | 8.4 | 37.3 | 6.1 | 0.1 | 10.2% | 16.3% |
| **Median** | 34 | 114.1 | 89.6 | 84 | 1943.4 | 1859 | 1502 | 21.1 | 10.7 | 22 | 39.8 | 37.3 | 33.5 | 0.9 | 100% | 34.8% |

Among the three departments of *Chemical Engineering* (see Table 7) the department at the University of Patras presents the highest numbers in most evaluation indexes in relation to the other two similar departments. It is also remarkable that the number of Academics at the Department of Athens is almost the double in comparison to the other two departments. However, the values of all indexes are quite lower than those at the other two departments.

Table 8: Aggregate results of Civil Engineering departments

| | | Publications | | | Citations | | | h - index | | | i10-index | | | m-index | Res. Act. | GS pr. |
|---|---|---|---|---|---|---|---|---|---|---|---|---|---|---|---|---|
| University | No. | Mean | SD | Median | Mean | SD | Median | Mean | SD | Median | Mean | SD | Median | Mean | | |
| NTUA | 61 | **101.1** | 123 | **73** | **936.9** | 1270 | **637** | **13** | 8.6 | **13** | **20.6** | 24.4 | **15** | **0.7** | 42.6% | 59.0% |
| Patra | 33 | 64.9 | 54.6 | 51 | 924.4 | 1363.7 | 330 | 12.4 | 9.4 | 10 | 18.1 | 20.6 | 11 | 0.6 | 97.0% | 60.6% |
| Thrace | 47 | 48.7 | 58 | 31 | 352.1 | 498.1 | 177 | 7.9 | 5 | 8 | 8.3 | 9.7 | 7 | 0.4 | 46.8% | 21.3% |
| Thessaly | 24 | 54.8 | 39.7 | 45 | 577.2 | 1150.7 | 252.5 | 9.3 | 7.4 | 7.4 | 11.2 | 15.3 | 7 | 0.5 | 87.5% | 54.2% |
| Thessaloniki | 94 | 54.7 | 74.9 | 31 | 508 | 1437.5 | 177.5 | 8 | 7.5 | 6.5 | 10.1 | 21 | 5 | 0.4 | 37.2% | 26.6% |
| **Mean** | *51.8* | *64.8* | *70.1* | *46.2* | *659.7* | *1144* | *314.8* | *10.1* | *7.6* | *9* | *13.7* | *18.2* | *9* | *0.5* | *62.2%* | *44.3%* |
| **SD** | *24.6* | *18.9* | *84.5* | *15.5* | *232.9* | *1259.9* | *170.7* | *2.2* | *8* | *2.3* | *4.8* | *20.4* | *3.6* | *0.1* | *24.9%* | *16.9%* |
| **Median** | *47* | *54.8* | *58* | *45* | *577.2* | *1270* | *252.5* | *9.3* | *7.5* | *8* | *11.2* | *20.6* | *7* | *0.5* | *46.8%* | *54.2%* |

Among the Civil Engineering departments (Table 8), the department at the University of Athens is ranked first in all 5 evaluation indexes, while the department at the Democritus University of Thrace scores the lowest numbers. The department at the University of Patras has the highest rates to members who report their research work at the department's web site and maintain GS profile.

Table 9: Aggregate result of Electrical and Computer Engineering departments

| | | Publications | | | Citations | | | h - index | | | i10-index | | | m-index | Res. Act. | GS pr. |
|---|---|---|---|---|---|---|---|---|---|---|---|---|---|---|---|---|
| University | No. | Mean | SD | Median | Mean | SD | Median | Mean | SD | Median | Mean | SD | Median | Mean | | |
| NTUA | 82 | **165** | 123.2 | **135** | 1912.1 | 2342.7 | 1126.5 | 18.7 | 10 | **17** | **39.6** | 34.9 | **29** | 0.8 | 47.6% | 47.6% |
| Crete | 28 | 96.3 | 60.2 | 90.5 | 1894.3 | 2020.2 | **1128.5** | 17.1 | 9.2 | 16 | 26.7 | 20.8 | 22.5 | **1** | 89.3% | 53.6% |
| Thessaly | 23 | 97.8 | 99.8 | 70 | **2153.1** | 4031.2 | 689 | **19** | 15.8 | 14 | 27.4 | 31.3 | 17 | **1** | 82.6% | 47.8% |
| Patra | 52 | 101.3 | 74 | 82.5 | 949.1 | 929.3 | 555.5 | 13.3 | 6.8 | 12 | 21.3 | 19.6 | 15 | 0.5 | **96.2%** | **92.3%** |
| Thessaloniki | 51 | 91.1 | 69.4 | 89 | 1037.2 | 1258.5 | 578 | 14 | 8.4 | 12 | 22.5 | 21.5 | 15 | 0.6 | 68.6% | 41.2% |
| Thrace | 48 | 83.3 | 55.4 | 62.5 | 809.4 | 758.4 | 512 | 12.9 | 6 | 11 | 19.4 | 16.2 | 12 | 0.6 | 75% | 37.5% |

| | | | | | | | | | | | | | | | |
|---|---|---|---|---|---|---|---|---|---|---|---|---|---|---|---|
| **Mean** | *47.3* | *105.8* | *80.4* | *88.3* | *1459.2* | *1890* | *764.9* | *15.8* | *9.4* | *13.7* | *26.1* | *24* | *18.4* | *0.8* | *76.5%* | *53.3%* |
| **SD** | *19.2* | *27.1* | *95* | *23.2* | *538* | *2029.7* | *261.9* | *2.5* | *9.5* | *2.2* | *6.7* | *27.1* | *5.7* | *0.2* | *15.8%* | *18.2%* |
| **Median** | *49.5* | *97* | *71.7* | *85.8* | *1465.7* | *1639.3* | *633.5* | *15.5* | *8.8* | *13* | *24.6* | *21.1* | *16* | *0.7* | *78.8%* | *47.7%* |

From Table 9, it emerges that the department at the National Technical University of Athens has the highest values in Publications. As far as Citations and h-index are concerned, the departments at the NTUA, Crete and Thessaly present a quite similar performance. The department of Thrace seems to have the lowest numbers in almost all indices (except m-index) and in GS profile possession as well. The department at the university of Patras have the highest rates of GS profile use and personal webpages reporting academic activity among their members.

Table 10: Aggregate results of Mechanical Engineering departments

| DEPARTMENTS OF MECHANICAL ENGINEERING | | | | | | | | | | | | | | | | |
|---|---|---|---|---|---|---|---|---|---|---|---|---|---|---|---|---|
| | | Publications | | | Citations | | | h - index | | | i10-index | | | m-index | Res. Act. | GS pr |
| **University** | **No.** | **Mean** | **SD** | **Median** | **Mean** | **SD** | **Median** | **Mean** | **SD** | **Median** | **Mean** | **SD** | **Median** | **Mean** | | |
| Thessaly | 20 | 83.2 | 41.3 | 90.5 | 1244 | 1146 | **1007.5** | 17.1 | 7.6 | **18** | 26.3 | 17.2 | **27** | 0.7 | **95%** | 60% |
| Thessaloniki | 30 | **120** | 98 | **104** | 1519.8 | 1336.7 | 1003.5 | **17.2** | 9.2 | 16 | 32.2 | 29.8 | 24 | **0.8** | 33.3% | 30% |
| NTUA | 47 | 90.6 | 68.5 | 71 | 1177.5 | 1327 | 606 | 15.2 | 9.1 | 13 | 26.5 | 27.9 | 16 | 0.7 | 66% | 40.4% |
| Patra | 43 | 87.6 | 71.7 | 60 | 1021.3 | 1053.2 | 573 | 14.2 | 8.3 | 13 | 23.9 | 24 | 14 | 0.6 | 86% | **86%** |
| W. Maced. | 16 | 52.5 | 33 | 50.5 | 626 | 584.7 | 287.5 | 11.5 | 6.3 | 9 | 15.3 | 13.7 | 8 | **0.8** | 81.3% | 68.8% |
| **Mean** | *31.2* | *86.8* | *62.5* | *75.2* | *1117.7* | *1089.5* | *695.5* | *15* | *8.1* | *13.8* | *24.8* | *22.5* | *17.8* | *0.7* | *72.3%* | *57%* |
| **SD** | *12.2* | *21.5* | *73* | *19.6* | *294* | *1201.5* | *276.3* | *2.1* | *8.6* | *3.1* | *5.5* | *25.3* | *6.9* | *0.1* | *21.6%* | *20%* |
| **Median** | *30* | *87.6* | *68.5* | *71* | *1177.5* | *1146* | *606* | *15.2* | *8.3* | *13* | *26.3* | *24* | *16* | *0.7* | *81.3%* | *60%* |

The results in Table 10 show that the department of Mechanical Engineering at the University of Thessaly is the first in the ranking, while the department at the University of the Western Macedonia is the last. However, while the department located at the University of Western Macedonia shows low values at almost all the indices, has the highest m-index (indicating a department with relative young scholars). Moreover, despite the fact that the department at the University of Thessaloniki is ranked first in various indices such as h-index and i-10 index, has the lowest percentage of members who report scientific activity on the web site of the department. The departments at the universities of Patras and Thessaly have the highest rates of GS profile use and personal webpages reporting academic activity among their members.

Concluding, the biggest differentiations between departments of the same discipline are presented in the departments of Sciences. In some cases, the difference in the h index between the first and the last Department exceeds 11 points, and the difference in the mean number of publications 100. The deviations in Engineering Departmetns are rather smaller (up to 6 h-index points and 82 publications, respectively).

**RQ2. Differences between academics who report detailed information about their research on the department's website and those who don't**

Further analysis was conducted to examine possible link between scientific output and scholars' who report academic activity. The purpose was to determine whether there was any difference in the level of their research, among those who report their scientific activity and those who do not. In Table 11 the data collected for the departments of each academic discipline are presented.

Table 11: Differences between academics who report detailed information about their research on the department's website and those who don't (* indicates statistical significance at the .005 level)

| Departments | p-value | | |
|---|---|---|---|
| | Publications | Citations | h-index |
| Mathematics | **0,001*** | **0,001*** | **<0,001*** |
| Statistics | 0,242 | 0,513 | 0,281 |
| Physics | **0,002*** | **0,001*** | **<0,001*** |
| Chemistry | **0,032*** | 0,071 | 0,124 |
| Biology | 0,319 | 0,318 | 0,45 |
| Computer Science | 0,052 | **0,049*** | 0,075 |
| Chemical Engineering | **<0,001*** | **<0,001*** | **<0,001*** |
| Civil Engineering | **<0,001*** | **<0,001*** | **<0,001*** |
| Electrical and Computer Engineering | **0,014*** | **0,015*** | **0,003*** |
| Mechanical Engineering | 0,129 | 0,082 | 0,055 |

A statistically significant difference emerged in three evaluation indices, publications, citations and h-index (Mann Whitney U). The results showed that in 50% of the examined departments (25/50), there was a statistically significant difference in publications,h-index and citations among those who were reporting their scientific activity on the department's web site and those who did not. Also, in 24% of the departments (Biological, Statistical, mechanical engineers) the assessment indicators were not different between those who reported scientific activity on the department's web site and those who did not. On the other hand, in the departments of Chemistry statistically significant difference existed only in publications index and in sections of it there was a statistically significant difference only in citations index. It should be noted that the differences were of statistical significance in all indices in the departments of Mathematics, Physics, Civil Engineering, Chemical Engineering, Electrical and Computer Engineering.

**RQ3. Differences between academics due to the location in which they received their PhD (Greece, Europe, USA)**

Differentiations to the research performance of faculty members, according to the source of their PhD were examined. Towards this end the scholars were into the following three categories: Scholars who obtained their PhD from Greece (GR, N=1138), Europe (EU, N=410) and United States of America (USA, N=318), respectively. An Internet search was conducted to find the origin of the PhD. The national PhD theses archive was as an additional source of information ( http://www.didaktorika.gr/eadd ). After a thorough investigation, it has been possible to collect information for the majority of faculty members under consideration, namely 1875. The difficulties at this point, had to do (a) with the lack of detailed CV, (b) with non-inclusion of a doctoral dissertation at the national archive of PhD theses, or (c) lack of response to personal communication via email. Thus, for 103/1978 academics the region in which they earned their PhD was not identified and in turn they were excluded from the study. Other 9 scholars completed their doctorate studies in other countries (Russia 2, Israel 1,Japan 1, South Africa 1, Australia 2, Hong Kong 2 and they were also excluded from the sample.

Analysis of the data showed that in general, there were significant deviations in all indices, h-index (by using the non-parametric Kruskal-Wallis H test, $\chi2=69.045$, $p<.001$), publications (Kruskal-Wallis H test, $\chi2= 56.651$, $p<.001$) and citations (Kruskal-Wallis H test, $\chi2= 81.143$, $p<.001$), depending on the scholars' doctorate source. In addition, further analysis for each pair was conducted (i.e. GR-EU, GR-US, EU-US). The analysis showed that for the pairs GR-US and EU-US there were statistically significant differences in all indices (in favor of US Mann-Whitney U, h-index GR-US: U= 131088, $p<.001$, publications GR-US: U= 136303.5, $p<.001$, citations GR-US: U= 126538, $p<.001$, h-index EU-US: U= 52446, $p<.001$, publications EU-US:U= 126538, $p<.001$, citations EU-US: U= 50790.5, $p<.001$). For the pair GR-EU a statistically significant difference emerged only in the number of publications (in favor of the EU: Mann-Whitney U h-index GR-EU: U= 233948.5, $p= .089>.05$, publications U=228655.5, $p= .019<.05$, citations GR-EU: U= 232436.5, $p= .059 >.05$).

In specific, in 7/10 of departments' categories a statistically significant difference in all indicators of assessment emerged depending on the region where PhD was obtained. As mentioned above, further analysis for each pair was conducted (i.e. GR-EU, GR-US, EU-US). It seemed that differences were evident in the pair of GR-US, in which 76% (38/50) of the departments statistically significant differences were detected in indicators in favor of scholars who earned their PhD in the US. Less significant differentiations were evident in the other pairs: As far as the GR-EU pair is concerned, in only 12/50 of the departments were found significant differences in all indicators in favor of the scholars who obtained their PhD in a European country other than Greece. In the 40% of the examined departments a statistically significant difference was found at least to one index (publications, citations or h-index). As for the EU-USA pair, in only 20% of the examined departments a statistically significant difference at least in one indication was found.

**RQ4. Correlation between academic rank, h-index and number of citations**

The last research question was related to the investigation of the link between the members' academic rank and their h-index and number of citations. In Table 12 the data obtained for the departments of each academic discipline are presented. In general, higher correlation suggests better hiring practices, since a higher scientific production is required to achieve higher academic ranks.

Table 12: Correlations between academic rank h-index and department (* indicates statistical significance at the .005 level)

| Departments | Academic rank- h-index correlation | Academic rank- citations correlation |
|---|---|---|
| Biology | 0.61* | 0.52* |
| Mechanical Engineering | 0.60* | 0.48* |
| Chemical Engineering | 0.50* | 0.30* |
| Physics | 0.47* | 0.40* |
| Computer Science | 0.42* | 0.37* |
| Mathematics | 0.40* | 0.21* |
| Civil Engineering | 0.39* | 0.27* |
| Chemistry | 0.38* | 0.31* |
| Electrical and Computer Engineering | 0.35* | 0.30* |
| Statistics | 0.17 | 0.27 |

The results showed that a statistically significant correlation between members' academic rank and h-index in 54% (27/50) of the departments. The values of the correlation coefficient Spearman's r ranged from .00 (lack of relationship between rank and h-index, Statistics Department of Statistics Piraeus) to. 82 (very strong correlation between rank and h index, Department of Computer Science, University of Thessaly). In addition, in 28% (14/50) of the examined departments a significant correlation between members' academic rank and citations was found.

## 4. Conclusions

The purpose of this study was to evaluate faculty members' research performance in departments of Sciences and Engineering in Greece using scientometric indices. Using the Internet and the citation database Google Scholar as well Publish or Perish software (PoP, Harzing, 2010), indices such as publications (publications), references (citations), h-index, i10-index and m-index were collected. The process was quite efficient and accurate. In the majority of the evaluated departments, a significant difference in h-index was observed between academics who report scientific activity on the departments' website and those who do not. Moreover, academics who earned their PhD title in the USA have higher indices in comparison to scholars who obtained their PhD title in Europe or in Greece. Finally, the correlation between the academic rank and the scholars' h-index (or the number of their citations) in some cases is quite low in some departments which, in some cases, could indicate lack of meritocracy.

From the before mentioned discussion, it arises that one of the ways to assess the quality of Greek universities, is by focusing on the research output of faculty members. From the results obtained some useful conclusions were derived, which could contribute to the improvement of the departments. Moreover, such studies could inform elected officials and policy makers and better shape the public opinion. For instance, it would be advisable for potential university students to choose departments based on the reported level of research and not on criteria such as distance from their place of residence. Apparently, in many cases other parameters, such as socio-economic status of each family, are involved. Moreover, rankings based on quantitative and widely accepted criteria would help to shape incentives for further research by all the Institutions. Appropriate interventions and policies could also aid the Universities to reach satisfactory scientific output.

The implementation of bibliometric evaluation in Greek Universities on an annual basis, will enable them to self-monitor the progress of the scientific output and the degree to which each University qualifies and meets tangible objectives laid down by the Greek Ministry of Education. Research for faculty members of academic departments, is one of their basic obligations as faculty members (apart from teaching and administrative tasks) and one could classify them into scientifically active or inactive. In this way, with the appropriate reform impetus an improvement of the current state of the Greek educational system could occur. More specifically, the evaluation indicators used in this work (h-index, m-index, publications, citations, i10-index) could be taken into account, with new legislation, to further incorporate transparent and merit practices into Tertiary Education. For instance, in some U.S. universities and disciplines science teachers are requested to have an index at least 12 to be promoted to the rank of associate professor and h equals 18 or higher enables their promotion to the rank of full professor (Lazaridis, 2008, p. 75). Similar policies in our country will provide greater research incentives to the faculty members to further improve the quality of their research work. In addition, transparent hiring practices based on tangible criteria adopted by hiring committees could motivate young people to pursue careers in academia.

However, the study is not without limitations. Our data are reliant on websites to determine the academic ranks of scholars. Therefore, we may have been inaccurate in assigning academic ranks to some of the academics in the present study. Also in some cases a synonymity occurred thus inflating the scholar's indices. In general, in only a few cases such a problem was occurred, since a lot of academics retain Google Scholar Profile. The latter is considered as accurate, which in turn in some cases could not be the case.

It is quite evident that the research questions answered in this paper, cover only a fraction of the possibilities provided by the bibliometric evaluation and statistical analysis of research data of the faculty. Numerous research questions and evaluation indicators of academic performance can be thoroughly studied (Mingers, & Leydesdorff, 2015). Our research was mainly focused to indices such as h-index, publications and citations. In addition, questions related to the area in which the scholars' PhD is obtained and possible relation to their output, as well as tendencies such as maintenance of GS profile and reporting scientific activity on the departments' web site, were also investigated. Other issues such as academic inbreeding (Inanc & Tuncer, 2011), relation between state funding and performance, number and characteristics of doctorate students and their scientific output as well as relation between scholars' gender and salaries and performance could be closely monitored and explored. A useful extension of this work is to incorporate representative international departments in order to better monitor the scientific progress and examine whether it is calibrated to an international level.

Finally, in order to publicly provide more reliable and representative results data should be collected by all Greek Universities and departments, preferably by using a suitable, usable and accessible web application (Katsanos, Tselios, Tsakoumis & Avouris, 2012, Orfanou, Tselios & Katsanos, 2015). Thus, the data related to scientific output could be instantly available to any stakeholder without the need to further process them and gives the ability to inform decisions. Therefore, a more organized, comprehensive and official effort to evaluate all the departments of the universities in the country, could greatly assist improvement of Greek Tertiary Education.